\documentclass[
 amsmath,amssymb,
 aps,
 nofootinbib,
pra,reprint,superscriptaddress
]{revtex4-1}

\usepackage{graphicx}
\usepackage{dcolumn}
\usepackage{bm}
\usepackage{units}
\usepackage{siunitx}
\graphicspath{{./figures/}} 
\usepackage[]{amsmath,amsfonts,amssymb}
\usepackage{physics}
\usepackage{color}
\usepackage[table,xcdraw,dvipsnames]{xcolor}
\usepackage[symbol*]{footmisc}
\usepackage{commath}
\usepackage{makecell}
\usepackage{calc}
\usepackage[normalem]{ulem}

\newcommand{\fe}{$^{57}$Fe~}
\newcommand{\bb}{BabyIAXO~}
\newcommand{\bbc}{BabyIAXO}
\definecolor{darkblue}{rgb}{1, 0.1, 0.2}

\newcommand{\exclude}[1]{}

\def\beq{\begin{equation}}
\def\eeq{\end{equation}}

\usepackage{natbib}
\setcitestyle{sort&compress,numbers,square}

\begin{document}
\numberwithin{equation}{section}

\title{Probing the axion-nucleon coupling with the next generation of axion helioscopes}

\author{Luca Di Luzio}
\email{luca.diluzio@unipd.it}
\affiliation{Dipartimento di Fisica e Astronomia `G.~Galilei', Universit\`a di Padova, Italy}
\affiliation{Istituto Nazionale di Fisica Nucleare, Sezione di Padova, Italy}

\author{Javier~Galan}
\email{javier.galan@unizar.es}
\affiliation{Center for Astroparticles and High Energy Physics (CAPA), Universidad de Zaragoza, 50009 Zaragoza, Spain}

\author{Maurizio Giannotti}
\email{MGiannotti@barry.edu}
\affiliation{Physical Sciences, Barry University, 11300 NE 2nd Ave., Miami Shores, FL 33161, USA}

\author{Igor~G.~Irastorza}
\email{Igor.Irastorza@cern.ch}
\affiliation{Center for Astroparticles and High Energy Physics (CAPA), Universidad de Zaragoza, 50009 Zaragoza, Spain}

\author{Joerg Jaeckel}
\email{jjaeckel@thphys.uni-heidelberg.de}
\affiliation{Institut f\"ur theoretische Physik, Universit\"at Heidelberg,
Philosophenweg 16, 69120 Heidelberg, Germany}

\author{Axel Lindner}
\email{axel.lindner@desy.de}
\affiliation{Deutsches Elektronen-Synchrotron DESY, Notkestr. 85, 22607 Hamburg, Germany}

\author{Jaime Ruz}
\email{ruzarmendari1@llnl.gov}
\affiliation{Lawrence Livermore National Laboratory,
7000 East Avenue, Livermore, CA 94551}

\author{Uwe Schneekloth}
\email{uwe.schneekloth@desy.de}
\affiliation{Deutsches Elektronen-Synchrotron DESY, Notkestr. 85, 22607 Hamburg, Germany}

\author{Lukas Sohl}
\email{lukas.sohl@desy.de}
\affiliation{Deutsches Elektronen-Synchrotron DESY, Notkestr. 85, 22607 Hamburg, Germany}

\author{Lennert J. Thormaehlen }
\email{l.thormaehlen@thphys.uni-heidelberg.de}
\affiliation{Institut f\"ur theoretische Physik, Universit\"at Heidelberg,
Philosophenweg 16, 69120 Heidelberg, Germany}

\author{Julia K. Vogel}
\email{vogel9@llnl.gov}
\affiliation{Lawrence Livermore National Laboratory,
7000 East Avenue, Livermore, CA 94551}

\begin{abstract}
\noindent
A finite axion-nucleon coupling, nearly unavoidable for QCD axions, leads to the production of axions via the thermal excitation and subsequent de-excitation of \fe isotopes in the sun. 
We revise the solar bound on this flux adopting the up to date emission rate, and investigate the sensitivity of the proposed International Axion Observatory IAXO and its intermediate stage BabyIAXO 
to detect these axions.
We compare different realistic experimental options and discuss the model dependence of the signal. Already \bb has sensitivity far beyond previous solar axion searches via the nucleon coupling and IAXO can improve on this by more than an order of magnitude.
\end{abstract}

\maketitle

\section{Introduction}
\label{sec:Intro}

Axions~\cite{Weinberg:1977ma,Wilczek:1977pj} are the direct and testable prediction following from the Peccei-Quinn (PQ) mechanism~\cite{Peccei:1977hh,Peccei:1977ur} proposed as a possible solution to the \textit{strong CP problem} of the Standard Model (SM) of particle physics. 
This problem, still one of the most puzzling in modern particle physics, concerns the apparent absence of CP-violating effects in Quantum Chromodynamics (QCD).
To emphasize their role in explaining the observed QCD behavior, 
such particles are also dubbed \emph{QCD axions}. 
Other kinds of pseudoscalar particles, with properties very similar to the QCD axion but no relation to the strong CP problem, emerge naturally in several extensions of the SM, in particular in compactified string theories~\cite{Witten:1984dg,Conlon:2006tq,Arvanitaki:2009fg,Acharya:2010zx, Higaki:2011me,Cicoli:2012sz,Demirtas:2018akl,Mehta:2021pwf}. 
To distinguish them from the QCD axion, these are often called axion-like particles (ALPs). 
The considerations discussed in this paper apply, in large part, to the QCD axion as well as to ALPs, except when specific QCD axion models are discussed. 
In the interest of brevity, we will refer to both QCD axions and ALPs as ``axions'' throughout this paper.

The axion phenomenology and experimental landscape are discussed in several recent  reviews~\cite{Irastorza:2018dyq,DiVecchia:2019ejf,DiLuzio:2020wdo,Agrawal:2021dbo,Sikivie:2020zpn}. 
In summary, axions are expected to couple to the SM fields with model dependent couplings,
\begin{equation} 
\label{eq:L_int}
\mathcal{L}_{\rm int} = -\frac{1}{4} g_{a\gamma} a F_{\mu\nu} \tilde F^{\mu\nu}
- \sum_f i\,g_{af}\, a\,\bar f \gamma_5 f \, ,
\end{equation}
where $F$ is the electromagnetic field tensor and $f$ are the SM fermions (for the present work, it is sufficient to consider couplings to protons, neutrons and electrons, so we can assume $f=p,n,e$).

Particularly appealing is the axion coupling to photons, since it allows very promising experimental strategies for the axion search. 
However, the couplings to electrons and nucleons are also employed in experimental axion searches.

In this work, we examine the axion coupling to nucleons (for a recent calculation see~\cite{GrillidiCortona:2015jxo}) and explore the potential to probe it with axion helioscopes~\cite{Sikivie:1983ip} which, as suggested by the name, search for axions produced in the sun. 

From a theoretical point of view the nucleon coupling is of particular interest because it receives, similar to the photon coupling, an unavoidable contribution from the defining coupling of QCD axions to gluons.
While it is possible to conceive models where this is (partially) cancelled, thereby having QCD axions with a suppressed coupling to nucleons~\cite{DiLuzio:2017ogq}, a large nuclear coupling is expected in most models which solve the strong CP problem and in certain cases can be considerably enhanced (see, e.g., Ref.~\cite{Darme:2020gyx}). 
A sizable nucleon coupling is therefore an expected feature of a QCD axion. 
In this sense, measuring the axion-nucleon coupling would be a good indication of the QCD axion nature.
Furthermore, such a detection
by a helioscope would likely be accompanied by a signal from Primakoff (and possibly Compton/Bremsstrahlung) axions. 
All these spectra have well defined shapes and so, given enough statistics and energy resolution, the data analysis might help understand 
specific properties of the detected particle
(see discussion in Ref.~\cite{Jaeckel:2018mbn}
for a  similar analysis in the case of the axion-photon and axion-electron couplings).

The axion coupling to nucleons can be probed indirectly in astrophysics, through the effects on the cooling of neutron stars (NS)~\cite{Keller:2012yr,Sedrakian:2015krq,Hamaguchi:2018oqw,Beznogov:2018fda,Sedrakian:2018kdm,Leinson:2021ety} and from the analysis of the observed neutrino signal from SN 1987A~\cite{Turner:1987by,Burrows:1988ah,Raffelt:1987yt,Raffelt:1990yz,Carenza:2019pxu,Carenza:2020cis,Fischer:2021jfm}. 
A direct detection is also possible, in principle, through experiments such as
CASPEr-gradient~\cite{Garcon:2017ixh} or ARIADNE~\cite{Arvanitaki:2014dfa}.\footnote{Note, however, that both experiments rely on important assumptions which strongly affect the predicted sensitivities.
The CASPEr detection potential depends on the axion primordial abundance, and it is strongly reduced if axions constitute only a
small fraction of the total dark matter in the galaxy.
The ARIADNE potential 
depends on the CP-violating axion-scalar coupling, 
which is presently very poorly known (see.\ e.g., Ref.~\cite{Bertolini:2020hjc}).}

Axions coupled to nucleons could also be produced in the sun, for example through the decay of excited nuclear states.
As pointed out already in Weinberg's seminal paper~\cite{Weinberg:1977ma}, being pseudoscalars, axions can be emitted in magnetic nuclear transitions.
The best known example, and the one adopted here, is the decay of the first excited state of $^{57}$Fe with the emission of a 14.4\,keV axion. Other nuclear transitions turn out to generate a substantially smaller axion flux (see also Appendix~\ref{app:lines}).

The search for these $^{57}$Fe axions has a long history.
The currently most powerful helioscope, the CERN Axion Solar Telescope (CAST)~\cite{CAST:2009jdc}, as well as CUORE~\cite{CUORE:2012ymr} and, more recently, XENON1T~\cite{XENON:2020rca}, have  
searched for axions produced in this transition and provided constraints on the axion-nucleon coupling. 

In this paper, we assess the potential of the next generation of axion helioscopes, \bb~\cite{BabyIAXO:2020mzw,Abeln:2020eft} 
and IAXO~\cite{Irastorza:2011gs,IAXO:2019mpb,Giannotti:2016drd}, to detect such axions. 
An important motivating factor is that with \bb the transition from concept to real experiment is imminent, as its  construction is expected to start in 
2022. 
We therefore believe it is of great importance to assess its potential for an axion detection in all possible channels.
Furthermore, the \bb and IAXO potential to probe this coupling are expected to be considerably superior to that of CAST~\cite{CAST:2009jdc}, allowing to probe large regions of additional axion parameter space. 
We provide a guide for the best setups required to maximize the efficiency to detect axions from $^{57}$Fe. 
We also use this opportunity to include new theoretical developments. 
In particular, the matrix elements for the relevant transition have recently been revised~\cite{Avignone:2017ylv}, and show a $\sim$\,30\,\% increase in the axion emission rate. 
In addition, we use a more recent solar model than previous publications.
Finally, we discuss the model dependence of the \fe signal 
and identify a class of QCD axion models 
which yield an enhanced signal 
compared to the standard 
KSVZ~\cite{Kim:1979if,Shifman:1979if} 
and 
DFSZ~\cite{Zhitnitsky:1980tq,Dine:1981rt} 
axion models.

The paper is organized as follows.
In section~\ref{sec:ALPs_from_nuclear_transitions}, we revisit the problem of the solar axion production in nuclear transitions, provide an updated expression for the flux from \fe 
and discuss its dependence on the specific  axion model;
in section~\ref{sec:stars}, we 
present a brief discussion of the current stellar bounds;
section~\ref{sec:sensitivity_estimates} is dedicated to the sensitivity estimates for Baby\-IAXO and IAXO, assuming different experimental setups;
finally, in section~\ref{sec:Conclusion} we provide our conclusions. 
A discussion of other nuclei that could contribute spectral lines to the solar axion flux can be found in Appendix~\ref{app:lines}.

\section{Axions from nuclear transitions in the sun}
\label{sec:ALPs_from_nuclear_transitions}

Axion helioscopes~\cite{Sikivie:1983ip} such as the planned International Axion Observatory (IAXO)~\cite{Irastorza:2011gs, BabyIAXO:2020mzw, IAXO:2019mpb,Giannotti:2016drd,Abeln:2020eft} are searching for very light (sub-eV) axions produced in the sun. 
Axions can be produced in the solar plasma by various processes. 
Helioscopes search primarily for axions originating from the Primakoff effect (cf, e.g.~\cite{Raffelt:1985nk,Raffelt:1987yu, Raffelt:2006cw,Carosi:2013rla,Ayala:2014pea,Giannotti:2015kwo,Giannotti:2017hny}) or from axion-electron interactions~\cite{Redondo:2013wwa,Hoof:2021mld,Jaeckel:2019xpa,Jaeckel:2018mbn}. 
This makes them sensitive to both the axion-photon coupling $g_{a\gamma}$ and to the product of photon and electron couplings $g_{a\gamma}g_{ae}$~\cite{Barth:2013sma,Jaeckel:2018mbn}. 

Axions can also be emitted in solar nuclear processes.
There are two noteworthy mechanisms for this: nuclear fusion and decays as well as thermal excitation and subsequent de-excitation of the nuclei of stable isotopes.  

Axions produced in nuclear fusion and nuclear decay processes typically carry an energy of the order of $\sim$\,MeV.
The axion flux from the $p+d\to$ $^{3}{\rm He}+a$\,(5.5\,MeV) reaction,
which provides one of the most intense axion fluxes from nuclear reactions, has been experimentally searched for by Borexino~\cite{Borexino:2012guz}. 
More recently, the SNO data~\cite{Bhusal:2020bvx} has also been scrutinized for traces of these 5.5\,MeV axions.
Helioscopes can, in principle, also be equipped with a 
$\gamma$-ray detector to look for such axions.
For example, the CAST helioscope installed a $\gamma$-ray calorimeter for some time to gain sensitivity to these high-energy axions~\cite{CAST:2009klq}.
However, in general all these analyses provided somewhat 
weak bounds on the axion-nucleon coupling, 
since the solar flux from nuclear reactions is 
not expected to be very large.
Indeed, Ref.~\cite{CAST:2009klq} estimated an axion flux of the order of $10^{10}  (g^{3}_{aN})^{2} /({\rm cm^2s})$ for the reaction mentioned above. 
This is more than 10 orders of magnitude smaller than the flux we will find below, Eq.~\eqref{eq:totalflux}.

A perhaps more promising direction is to look for low lying nuclear excitations of stable isotopes with a significant abundance inside the sun  that can be thermally excited.
Two candidates have been proposed for this in the past, $^{57}$Fe \cite{Moriyama:1995bz,Krcmar:1998xn} and  $^{83}$Kr \cite{Gavrilyuk:2014mch}.
The former has a first nuclear excitation energy $E^*$ of 14.4\,keV and the latter of 9.4\,keV. 
In the solar core, at temperatures $T\sim1.3$\,keV, these excited states have a small but non-vanishing occupation number that can be calculated from the Boltzmann distribution.
The amount of axions produced is proportional to the occupation number, the isotope abundance and the inverse lifetime of the excited state. 
By combining a list of possible elements and their nuclear transitions \cite{Roehlsberger:2004} with the solar abundances in \cite{Asplund}, 
it becomes clear that for IAXO it is the $^{57}$Fe transition that would produce the strongest signal (see Appendix~\ref{app:lines} for more details).  

\subsection{Effective axion coupling in the $^{57}$Fe nuclear transition}
\label{sec:models}
To describe the axion interactions with nuclei it is convenient to rewrite the relevant terms in the Lagrangian~\eqref{eq:L_int} as
\begin{align}
\label{eq:geffdef2}
\mathcal{L}_{aN} 
= - i a \bar N \gamma_5 \left( g^0_{aN} + g^3_{aN} \tau^3 \right) N \,.
\end{align}
Here, $N = (p, n)^T$ is the nucleon doublet, 
$g^0_{aN}$, $g^3_{aN}$ are the iso-scalar and iso-vector couplings, respectively, 
and $\tau^3$ is the Pauli matrix.

The axion-to-photon branching ratio for the decay rates
of the first excited state of $^{57}$Fe can then be expressed as~\cite{Haxton:1991pu,Avignone:1988bv}
\begin{align}
\label{eq:Gamma_Ratio_general}
\frac{\Gamma_a}{\Gamma_\gamma}=
\left( \frac{k_a}{k_\gamma} \right)^3
\frac{1}{2\pi\alpha}\frac{1}{1+\delta^2}
\left[ \frac{\beta \, g_{aN}^{0} + g_{aN}^{3}} 
{\left( \mu_0-\frac12 \right) \beta + \mu_3 -\eta} \right]^2
,
\end{align}
where $k_a,~k_\gamma$ are the axion and photon momenta,   
$\mu_0$ and $\mu_3$ are the isoscalar and isovector nuclear magnetic moments (expressed in nuclear magnetons), 
$\delta$ is the E2/M1 mixing ratio for the $^{57}$Fe nuclear transition, 
while $\beta$ and $\eta$ are constants dependent on the nuclear structure~\cite{Avignone:1988bv}. 
The constant $\delta\simeq 0.002$ can be safely neglected to the level of precision required in this work. 
The other constants were recently reevaluated in Ref.~\cite{Avignone:2017ylv}.
Assuming $(k_a/k_\gamma)\simeq 1$, which applies to ultrarelativistic axions, and adopting the most recent 
values in Ref.~\cite{Avignone:2017ylv} for the relevant nuclear constants, we find
\begin{align}
\label{eq:Gamma_Ratio_specific}
\frac{\Gamma_a}{\Gamma_\gamma}
     &=2.32
    \left( -1.31\, g_{aN}^0 + g_{aN}^3 \right)^2\\
    &=2.32    \left( 0.16\, g_{ap} +1.16\, g_{an} \right)^2\,,
    \label{eq:ratio_Avignone}
\end{align}
where the second line is expressed in terms of the more common axion couplings to neutrons, 
\begin{equation}
    g_{an} =  g_{aN}^0 - g_{aN}^3\;,
\end{equation} 
and to protons, 
\begin{equation}
g_{ap} = g_{aN}^0 + g_{aN}^3\;.
\end{equation}
From the form of  Eq.~\eqref{eq:ratio_Avignone}, 
it is convenient to define the effective nucleon coupling as\footnote{We stress that this combination is specific to the 14.4 keV transition in iron. However, as this is by far the most promising transition we use it in the following without a specific label.}
\begin{align}
\label{eq:g_N_eff}
g_{aN}^{{\rm eff}}=0.16\, g_{ap} +1.16\, g_{an}\,,
\end{align}
which is the coupling combination that controls the axion emission rate in this transition. 
Notice that the updated branching ratio is 27\,\% larger than the one 
found in Ref.~\cite{Haxton:1991pu} and used in the previous experimental analyses, including CAST~\cite{CAST:2009jdc}, CUORE~\cite{CUORE:2012ymr} and more recently XENON1T~\cite{XENON:2020rca}.
Furthermore, the relative importance of the coupling to protons has strengthened in this new analysis even though, as evident from Eq.~\eqref{eq:ratio_Avignone}, it is still considerably less relevant than the coupling to neutrons.

\subsection{Axion model dependence 
of the \fe transition rate}\label{subsec:modeldep}

Before proceeding with the experimental sensitivity, it is interesting to look at the axion emission rate for some 
benchmark axion models. 

Expressing the axion-nucleon 
couplings in terms of the 
dimensionless
axion-quark 
coefficients $c^0_q$ \cite{diCortona:2015ldu}
(defined via the Lagrangian term
$\frac{\partial_\mu a}{2 f_a} c^0_q 
\bar q \gamma^\mu \gamma_5 q$,
with $f_a$ the axion decay constant), Eq.~\eqref{eq:ratio_Avignone} can be cast as\footnote{In deriving this equation, we use the standard relation (cf., e.g.,~\cite{GrillidiCortona:2015jxo})
\begin{align}
    m_a\simeq 5.7 \, \mu {\rm eV}\times \left( \frac{10^{12} \, {\rm GeV}}{f_a} \right)
\end{align}
between the axion mass and decay constant.
}
\begin{align}
\label{eq:Gamma_Ratio_quarkcoupl}
&\frac{\Gamma_a}{\Gamma_\gamma}
     =5.81 \times 10^{-16}
    ( 
    1 
    +3.28 \, c^0_u
    -9.97 \, c^0_d
    +0.52 \, c^0_s \nonumber \\
    &
    +0.16 \, c^0_c
    +0.12 \, c^0_b
    +0.048 \, c^0_t
    )^2 
    \left( \frac{m_a}{1 \, \text{eV}} \right)^2 \,,
\end{align}
where $m_a$ is the axion mass. 
In the KSVZ model \cite{Kim:1979if,Shifman:1979if}
one has $c^0_q = 0$, yielding
\begin{equation}
\label{eq:Gamma_Ratio_KSVZ}
\left. \frac{\Gamma_a}{\Gamma_\gamma} \right|_{\rm KSVZ}
     =5.81 \times 10^{-16}
\left( \frac{m_a}{1 \, \text{eV}} \right)^2 \,, 
\end{equation}
while in 
the DFSZ 
model \cite{Zhitnitsky:1980tq,Dine:1981rt}, 
$c^0_{u,c,t} = \frac{1}{3} \cos^2\beta$ 
and $c^0_{d,s,b} = \frac{1}{3} \sin^2\beta$ 
corresponding to 
\begin{align}
\label{eq:Gamma_Ratio_DFSZ}
&\left.\frac{\Gamma_a}{\Gamma_\gamma}\right|_{\rm DFSZ}
     =5.81 \times 10^{-16} \\
    &\times ( 
    1 
    + 1.16 \cos^2\beta - 3.11 \sin^2\beta
    )^2 
    \left( \frac{m_a}{1 \, \text{eV}} \right)^2 \,,  
    \nonumber
\end{align}
with $\tan\beta$ 
defined in the perturbative domain $\in[0.25, 170]$ \cite{DiLuzio:2020wdo}.

Note that in KSVZ models the axion emission rate is accidentally suppressed as in these models the neutron coupling is very small. In DSFZ models it can get enhanced by 
up to a factor of $\sim$\,4 with respect to the KSVZ. 

Eq.~\eqref{eq:Gamma_Ratio_quarkcoupl} 
suggests that a strong enhancement of the 
axion emission rate can be achieved 
for ``down-philic'' axions, with 
$c^0_d \gg c^0_u$.
More generally this holds if 
a sizable cancellation with the 
model-independent factor normalized to 1
in Eq.~\eqref{eq:Gamma_Ratio_quarkcoupl} 
is avoided. 
This last possibility naturally happens in a class 
of non-universal 
DFSZ models with $c^0_u + c^0_d = 1$ 
that have been analyzed in 
Refs.~\cite{DiLuzio:2017ogq,Bjorkeroth:2018ipq,Bjorkeroth:2019jtx} 
(for a summary of axion couplings in those 
models see Table~3 in \cite{DiLuzio:2021ysg}).
For instance, in the non-universal M1 model 
of Ref.~\cite{DiLuzio:2017ogq} 
one has 
$c^0_u = c^0_c = \sin^2\beta$, 
$c^0_t = - \cos^2\beta$, 
$c^0_d = c^0_s = \cos^2\beta$, 
$c^0_b = - \sin^2\beta$, 
with
$\tan\beta \in [0.25, 170]$
(also, $c^0_e = - \sin^2\beta$ and $E/N = 2/3$), 
leading to 
\begin{align}
\label{eq:Gamma_Ratio_M1}
&\left.\frac{\Gamma_a}{\Gamma_\gamma}\right|_{\rm M1}
     =5.81 \times 10^{-16} \\
    &\times ( 
    1 
        + 3.32 \sin^2\beta - 9.50 \cos^2\beta
    )^2 
    \left( \frac{m_a}{1 \, \text{eV}} \right)^2 \,,  
    \nonumber
\end{align}
which at small $\beta$
yields an $\mathcal{O}(60)$
enhancement 
of the axion emission rate 
with respect to the KSVZ model.\footnote{Again keeping $\beta$ in the perturbative range.} 
Other non-universal DFSZ models, 
among those mentioned above, 
feature a similar enhancement 
of the $g^{\rm eff}_{aN}$ 
coupling, but they have different 
values for the axion-photon coupling 
(that is important 
in detection for the IAXO setup).
In particular, 
the non-universal 
model $\mathcal{T}_2^{(u)}$ 
of Ref.~\cite{Bjorkeroth:2018ipq} 
features the largest 
axion coupling to photons 
among the general 
class of non-universal DFSZ models with two Higgs doublets 
(see Table 5 in \cite{DiLuzio:2021ysg}).

\subsection{Solar $^{57}$Fe axion flux and limits}
\label{sec:axion_flux}

To calculate the resulting flux from thermal excitation and subsequent de-excitation in detail, 
we follow the derivation by Moriyama \cite{Moriyama:1995bz}, also used by the CAST collaboration in their search for this source \cite{CAST:2009jdc},
but adopt updated nuclear matrix elements derived in Ref.~\cite{Avignone:2017ylv}. 
The axion emission rate per unit mass of solar matter is given by
\begin{equation}
\label{eq:axionspersolarmatter}
\mathcal{N}_a=\mathcal{N}\,\omega_1(T)\, \frac{1}{\tau_0}\frac{1}{1+\alpha} \frac{\Gamma_a}{\Gamma_\gamma},
\end{equation}
where $\mathcal{N}$ is the $^{57}$Fe number density per solar mass, $\omega_1$ the occupation number of the first excited state, $\tau_0$ the lifetime of the excited state, $\alpha$ the internal conversion coefficient and $\frac{\Gamma_a}{\Gamma_\gamma}$ the branching ratio of axion to photon emission, given in Eq.~\eqref{eq:ratio_Avignone}.
A detailed description of these parameters is given in Appendix~\ref{app:lines}.

\begin{figure}
    \centering
    \includegraphics[width=1\linewidth]{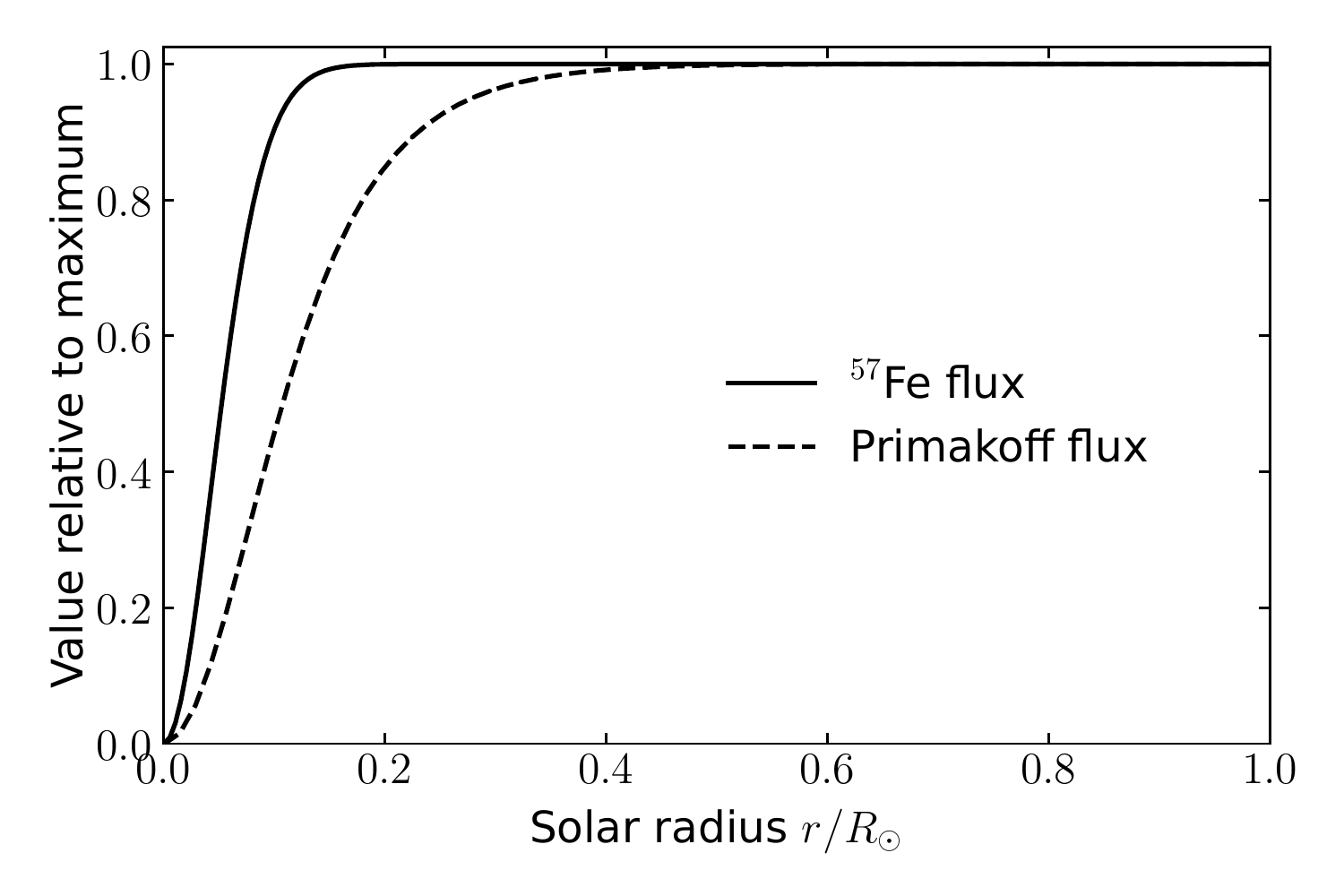}
    \caption{Radial dependence of the solar axion flux from $^{57}$Fe transitions and the Primakoff effect. We show the fraction of the flux inside the field of view which is a circle around the solar center with radius $r$. }
    \label{fig:flux-vs-radius}
\end{figure}

The axion signal from this nuclear transition is expected to be very narrow. 
The natural line width is negligible compared to the Doppler broadening $\sigma$ of 
\begin{equation}
\sigma(T)=E^*\sqrt{\frac{T}{m_{\text{Fe57}}}}\sim 2 \ \text{eV}\,,
\label{eq:width}
\end{equation}
corresponding to a full width at half maximum (FWHM) of $2.35\,\sigma \sim 5$ eV.
Consequently, the spectral axion flux at a distance of one astronomical unit $d_\odot$ from the sun is an integral of a Gaussian peak over the solar radius $R_\odot$,\footnote{We use the solar axion flux library~\cite{Hoof:2021mld}  which is publicly available at \url{https://github.com/sebhoof/SolarAxionFlux} to perform all axion flux calculations.}
\begin{align}
\frac{\dif\Phi_a(E_a)}{\dif E_a}=&\frac{1}{4\pi d_\odot^2} \int_{0}^{R_\odot} \mathcal{N}_a(T(r)) \frac{1}{\sqrt{2\pi}\sigma(T(r))} \nonumber \\
&\times \exp \left(-\frac{(E_a-E^*)^2}{2\sigma(T(r))^2}\right)\rho(r)4\pi r^2 \text{d}r.
\end{align}
Although, depending on the optics adopted, IAXO's field of view may cover only part of the sun,
we integrate over the entire solar radius since -- as 
illustrated in Fig~\ref{fig:flux-vs-radius} -- more than 99\,\% of the total flux originates from a circle around the solar centre with radius 0.15\,$R_\odot$.
The solar model (in our case B16-AGSS09~\cite{Vinyoles:2016djt}) fixes temperature $T$, density $\rho$ and iron abundance at every radius. 

It is noteworthy that this axion source is particularly sensitive to the temperature. This is because of the thermal occupation number $\omega_1 \propto e^{-E^*/T}$ of the excited nuclear state. For instance, the latest high and low metallicity solar models (B16-GS98 and B16-AGSS09 \cite{Vinyoles:2016djt}) only differ by about 1\,\% in their respective core temperature but this alone results in a difference of 12\,\% in the total flux.

We do not expect the narrow, Doppler-broadened peak to be resolved.\footnote{Metallic magnetic calorimeters~\cite{Unger2021} may reach an energy resolution of few eV which would enable them to coarsely resolve the Doppler peak.} Hence, only the total flux is of interest and the integral over $E_a$ can be performed to get the total solar axion flux from the $^{57}$Fe nuclear transition \cite{CAST:2009jdc},
\begin{align}
\label{eq:totalflux}
\Phi_a=&\frac{1}{4\pi d_\odot^2} \mathcal{N}
\frac{1}{\tau_0}\frac{1}{1+\alpha}
\frac{\Gamma_a}{\Gamma_\gamma}\nonumber \\&\times \int_{0}^{R_\odot} \omega_1(T(r))\rho(r)4\pi r^2 \text{d}r\\
=& 5.06 \times 10^{23}\ (g_{aN}^{{\rm eff}})^2 \ \rm{cm}^{-2}\rm{s}^{-1} \ .
\end{align}
This numerical result is larger than the one previously derived by the CAST collaboration~\cite{CAST:2009jdc} because we worked with the updated nuclear matrix elements. In addition, we integrated over a more recent solar model, namely B16-AGSS09~\cite{Vinyoles:2016djt}, whose core temperature and iron abundance is smaller compared to the values adopted previously. This explains why the overall flux is not equally enhanced as the updated nuclear matrix elements. 
The following calculations are all done with this updated axion flux and the cited bounds derived by the CAST collaboration in~\cite{CAST:2009jdc} were rescaled accordingly.

\bigskip

\section{Confronting stellar bounds}
\label{sec:stars}

Let us start by quickly updating the energy loss constraint.
Using the revised axion rate, Eq~\eqref{eq:ratio_Avignone}, and the solar model B16-AGSS09 we find a total axion luminosity (energy loss rate), 
\begin{equation}
   L_a=  8.38\times 10^9 (g_{aN}^{{\rm eff}})^2\,L_{\odot} 
\end{equation}
via the $^{57}$Fe transition.
The recent study in Ref.~\cite{Vinyoles:2015aba} constrains an exotic energy loss to a maximum of 3\,\% of the standard solar luminosity $L_{\odot}$.
This leads to an improved bound on the axion effective coupling with nucleons, 
\begin{equation}
g_{aN}^{{\rm eff}}\leq 1.89\times 10^{-6}\,. 
\label{eq:solar_bound}
\end{equation}
Notice that the result in Eq.~\eqref{eq:solar_bound}, known as the \emph{solar bound} on $g_{aN}^{{\rm eff}}$,
is about a factor of two more stringent than the 
previous constraint, $g_{aN}^{{\rm eff}}\leq 3.6 \times 10^{-6}$~\cite{CAST:2009jdc}. 
Besides the enhanced emission rate~\cite{Avignone:2017ylv} and the updated solar model~\cite{Vinyoles:2016djt}, this is also due to Ref.~\cite{CAST:2009jdc} having  excluded only $L_a > 0.1 \ L_\odot$.

\bigskip

The axion-nuclear coupling can also be constrained from other stellar observations. 
In particular, strong bounds were derived from X-ray observations of various NS~\cite{Keller:2012yr,Sedrakian:2015krq,Hamaguchi:2018oqw,Beznogov:2018fda,Sedrakian:2018kdm,Leinson:2021ety}.
These bounds are, however, subject to several uncertainties
and do not always agree with each other, not even when referring to the same star~\cite{DiLuzio:2021ysg}.
In any case, all these analyses suggest 
a limit 
of $\sim 10^{-9}$ on some combination of axion-nucleon couplings.\footnote{In most of these analyses, the limit applies only to the axion-neutron coupling.}

\begin{figure}
    \centering
    \includegraphics[width=1\linewidth]{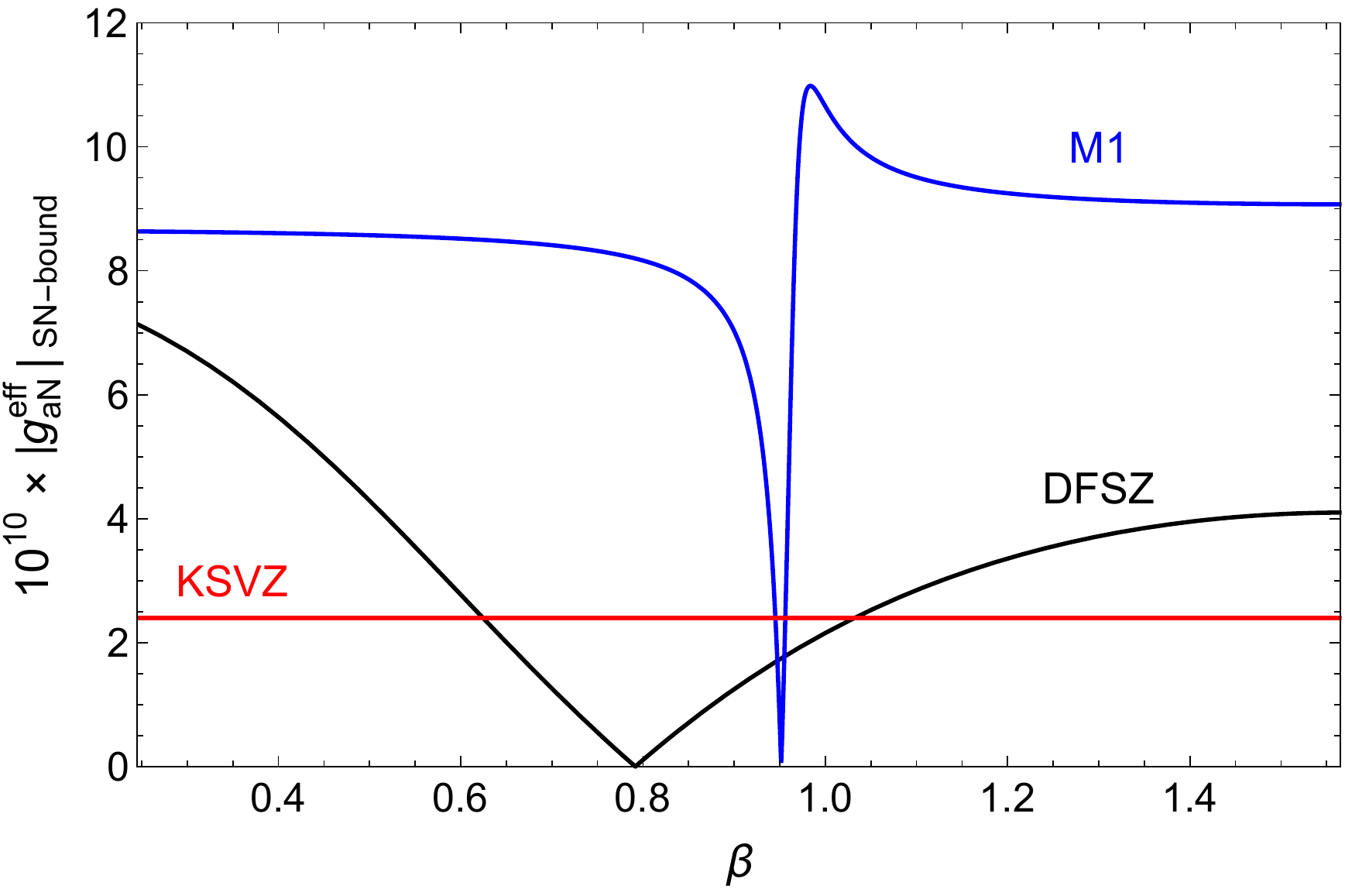}
    \caption{
    Effective axion coupling entering in the $^{57}$Fe transition as a function of the angle $\beta$, which defines the model dependent couplings (Cf.~Sec.~\ref{subsec:modeldep}).
    In the figure, the colored lines indicate the value of the coupling $g_{aN}^{\rm eff}$
    calculated assuming that the specific axion model saturates the SN bound, given in Eq.~\eqref{eq:gan_gap_SN_bound}.
}
    \label{fig:models_vs_SN}
\end{figure}

A similar bound can be deduced from the analysis of the neutrino signal observed in coincidence with the SN 1987A event~\cite{Raffelt:1987yt,Turner:1987by,Raffelt:1990yz,Turner:1991ax,
Raffelt:1993ix,Keil:1996ju,Hanhart:2000ae,Fischer:2016cyd,Carenza:2019pxu,Fischer:2021jfm}. 
Here, we refer specifically to the most recent analysis~\cite{Carenza:2019pxu}, which derived the bound
\begin{equation} 
g_{an}^2+ 0.6\, g_{ap}^2 + 0.5\, g_{an}\,g_{ap}\lesssim 8.3 \times 10^{-19} \,.
\label{eq:gan_gap_SN_bound}
\end{equation}
This is a very strong constraint and, as we will see, only advanced setups may allow the exploration of smaller couplings.
It is worth noticing, however, 
that the 
coupling expected in the $^{57}$Fe transition differs from 
the coupling in Eq.~\eqref{eq:gan_gap_SN_bound} and that, for some
specific models, the coupling relevant for the solar axion searches may be enhanced or suppressed with respect to what is constrained by the supernova (SN) argument (see Fig.~\ref{fig:models_vs_SN}).
For now, we will not investigate this argument further but
we stress that, due to the peculiar effective coupling 
appearing in the $^{57}$Fe transition,
the comparison with the astrophysical bounds is model dependent.

That said, Eq.~\eqref{eq:gan_gap_SN_bound} can be translated into a limit on $g^{\rm eff}_{aN}$ by choosing the ratio between the proton and neutron couplings such that the left hand side of this equation is minimal, $g_{ap}/g_{an}\approx -0.2$ while keeping $g^{\rm eff}_{aN}$ constant.
This yields,
\begin{equation}
\label{eq:SNbound}
  \left|g^{\rm eff}_{aN}\right|\lesssim 1.1\times 10^{-9}.
\end{equation}
The only model in Fig.~\ref{fig:models_vs_SN} which saturates this bound is the M1 model.

\section{Sensitivity estimates}
\label{sec:sensitivity_estimates}

In this section, we present a detailed discussion of the helioscopes potential to detect the $^{57}$Fe axion flux.
A particular emphasis is given to \bbc, 
which is expected to 
start construction soon at DESY in Hamburg.
Estimates for the more advanced IAXO and IAXO+ configurations~\cite{IAXO:2019mpb} will also be presented.

Before going into specific experimental aspects let us make a few general considerations that can provide some help with the experimental design.
To detect the $^{57}$Fe line we want to maximize the signal to noise ratio in the single relevant energy bin which contains all of the signal events. 
There are two main contributions to the background for the measurement of the nucleon line. 
One is the usual background rate of the detector itself (e.g., cosmic rays, environmental gammas, intrinsic detector radioactivity) that usually grows linearly with the area of the detector and it is typically proportional to the  spectral size of the signal bin. 
This means that, in the case of the expected narrow signals, it can be reduced by making use of good energy resolution detectors. Second, there is the physics background due to Primakoff production. This grows with the photon coupling. Importantly, as the Primakoff spectrum is continuous, it also grows linearly with worsening energy resolution. 
Combining these two effects leads us to the following figure of merit,
\begin{equation}
\label{eq:merit}
    f\propto \frac{S}{\sqrt{B}}\propto \frac{\epsilon_{o}\epsilon_{d}\ g^2_{a\gamma}}{\sqrt{\Delta E_d}\sqrt{b a+ g^{4}_{a\gamma}\kappa \epsilon_{o}\epsilon_{d}}}.
\end{equation}
Here, $\epsilon_{o,d}$ are the optics and detector efficiencies, $\Delta E_d$ is the energy resolution 
of the detector\footnote{In the numerical calculations in Sects.~\ref{sec:massless} and \ref{sec:sensitivity_massive} the detector resolution is implemented by a sharp signal bin of width  $\Delta E_d$ centered around the $^{57}$Fe peak.}, $b$ is the (spectral) background rate 
per area, and $a$ is the signal spot area on the detector. $\kappa$ quantifies the Primakoff flux in the $^{57}$Fe signal bin and is implicitly defined in Eq.~\eqref{eq:kappa}.
From this we can identify the parameters to be optimized. 
While good energy resolution is critical this should not be offset by too large a background. Similarly, with focusing optics we can reduce the detector area and therefore the background contribution. However, again there is a balance because such X-ray optics may have an efficiency significantly smaller than $1$. We will see all this more concretely when considering explicit setups below.

\subsection{\bbc}
\label{sec:BabyIAXO}
Conceived as an intermediate stage towards the full IAXO experiment, \bbc~\cite{BabyIAXO:2020mzw,Abeln:2020eft} is nevertheless expected to  substantially advance the exploration of the axion parameter space.
With its sensitivity, the experiment will be able to study QCD axion models, and to investigate stellar cooling hints and other well motivated sections of the parameter space~\cite{IAXO:2019mpb,DiLuzio:2020wdo}. 

The \bb experiment is mainly designed to measure Primakoff axions in the energy range from 1\,keV to 10\,keV with the peak of the solar axion flux spectrum at $\sim$3\,keV.
\bb will consist of two magnetic bores of 10\,m length and 70\,cm diameter, each with an average magnetic field strength of about 2\,T. 
Together with newly developed X-ray optics and detector systems providing higher energy resolution and lower background, \bb will be the first helioscope to exceed the sensitivity of the CAST experiment.
The magnet bores of \bb are of similar diameter as those of IAXO (60 cm) and IAXO+ (80 cm), 
thus the experience from the optics and detector development for \bb can later be applied to IAXO \cite{BabyIAXO:2020mzw}.
The first detection system for BabyIAXO is chosen to be microbulk micromegas technology. This detector technology has proven background levels as low as 
$10^{-7}/(\textrm{keV}\, \textrm{cm}^2\, \textrm{s})$ \,\cite{Aune_2014}
and a high detection efficiency for Primakoff photons $<$10\,keV. 
A variety of other detector types, like silicon drift detectors (SDD), metallic magnetic calorimeters (MMC) and transition edge sensors (TES), are also studied for \bb aiming to optimize the energy resolution for precision measurements of the axion spectrum\,\cite{BabyIAXO:2020mzw}.

\subsection{\bbc~configurations for $^{57}$Fe~detection}

For the measurement of $^{57}$Fe axions at 14.4\,keV, the detection efficiency in the baseline \bb configuration is not optimal 
and an enhancement of the 
detection system is necessary.
Indeed, GEANT4 simulations of the Micromegas detector to be used in the baseline \bb configuration
show that the ionization probability of 14.4\,keV photons at a conversion length of 3\,cm at the given Argon gas mixture and pressure is only around~15\,\%.
Additionally, the current designs of the \bbc~telescopes are not optimized for energies above $10$~keV~\cite{Armengaud:2014gea}. 
The expected optics efficiency at 14.4\,keV is a mere $1.3\,\%$ (cf.~``baseline''/\bbc$_0$ configuration in Tab.~\ref{tab:setups}).

In the following we discuss different adaptations to the detector system of \bb for the measurement of 14.4\,keV photons with enhanced efficiency, energy resolution and lower background.

Let us start with a relatively minimal modification.  
The sensitivity of the Micromegas detector for 14.4\,keV photons can be enhanced by changing the gas mixture and adjusting the pressure. Gas mixtures with an inert gas of higher Z-value, like Xenon, show a higher conversion efficiency for 14.4\,keV photons, compared to Argon.
A suitable high ionization probability for 14.4\,keV photons of $>$90\,\% should be reachable with a 10\,cm photon conversion length and a Xenon based gas mixture at atmospheric pressure. 
Currently, microbulk Micromegas~\cite{microbulk:2010} are developed to have a square active area of 25\,cm~x~25\,cm~\cite{Castel:2019ykn} and the whole bore opening can be covered with a few detector tiles. 
As \bb features two bores, one could be used for operation without X-ray optics and this detector could be operated in parallel on the second magnetic bore. 
Removing the optics and covering of the whole area improves the detection efficiency, as this avoids the losses from inefficient X-ray optics at this energy.
However, the detector will have a higher background
due to the larger conversion volume
than the smaller micromegas.
This ``no optics'' configuration is denoted as ``no optics''/\bbc$_1$ in Tab.~\ref{tab:setups} and in the figures.

The second detection concept for 14.4\,keV photons at \bb is based on silicon drift detectors (SDD). This detector consists of a thick negative doped layer, that is fully depleted by a negative bias voltage, and positive doped contacts and strips on both sides of the layer. The incoming X-ray radiation generates electrons in the depleted zone that are then drifted towards the anode at the end of the layer. 
An SDD detector of 300\,µm thickness may reach 50\,\% sensitivity for 15\,keV photons~\cite{sdd:2003}. This detector type is also considered as an additional detector for the baseline measurements at \bb \cite{BabyIAXO:2020mzw}. Efficient X-ray optics focusing the photons on the small detector structure are mandatory, as SDDs
are only produced in small pixels with a detector size in the range of millimeters.
Preliminary studies showed that the optimization of the X-ray optical system for high energy X-rays is possible,
using multilayer coating techniques~\cite{nustarcoat:2011}. 
The Nuclear Spectroscopic Telescope Array (NuSTAR)~\cite{nustar:2013} has already used X-ray optics sensitive to photons in the range of 5-80\,keV. 
A dedicated instrument with non-imaging optics can further boost the throughput at 14.4\,keV. 
A realistic figure for the optics efficiency is 
$\epsilon_0=0.3$, which is the value adopted in our analysis. 
In our study here, we assume that the optimized optics will be used in just one of the two \bb bores.
This assumption will be lifted in our analysis of IAXO and IAXO+.
With this included we obtain the ``optimized optics''/\bbc$_2$ configuration (cf.~Tab.~\ref{tab:setups}).

Finally, a third detection concept suitable for the detection of 14.4\,keV photons at \bb is based on Cadmium-Zinc-Telluride (CZT) semiconductors. 
The detection principle is similar to silicon-based ionization detectors.
With a higher Z-value compared to silicon, CZT provides a higher ionization probability for the photons and CZT of only 300\,µm thickness have an ionization probability for 14.4\,keV of $>$99\,\%.
Indeed CZT detectors are already used in experiments focusing on the detection of hard X-rays like NuSTAR~\cite{nustar:2013}.
These detectors can reach an energy resolution down to 2\,\% in the relevant energy range \cite{czt_eres:2005}, 
which makes them optimal for 
discrimination of the Primakoff background. However, the good energy resolution comes with the price of a somewhat increased background.
Furthermore, currently it is only possible to produce detectors with an active area of up to 2\,cm~x~2\,cm~\cite{czt:2005}, which again makes an efficient X-ray optic necessary for the use of CZT detectors at \bbc. 
Coupled with the same NuSTAR-like optics adopted in \bbc$_2$, this is our ``energy resolution''/\bbc$_3$ configuration.
Notice that, just like in the case of  \bbc$_2$, the detector and optics discussed above will be implemented in only  one of the two \bb bores (see Tab.~\ref{tab:setups}).

\subsection{IAXO and IAXO+}
The full scale helioscope IAXO 
is expected to begin construction during the \bb data-taking period. 
It adopts realistic components which will allow a considerably better performance than \bb in the entire mass range.  
IAXO+ is a more aggressive setup which will allow to
increase the sensitivity and a deep exploration of physically motivated axion parameters (see. e.g., Refs.~\cite{IAXO:2019mpb,DiLuzio:2021ysg}). 

Here, we consider two possible setups for IAXO/IAXO+, summarized in table~\ref{tab:setups}.
The first, indicated as IAXO$_\textrm{b}$,
uses the benchmark configuration parameters
discussed in the most recent IAXO publication~\cite{IAXO:2019mpb}.
In this case, we assume an energy resolution of 2\,\% in the 
relevant energy range, which is a very realistic figure for CZT detectors, as discussed in Sec.~\ref{sec:BabyIAXO}.
In addition we assume, somewhat optimistically, that significant improvements in the backgrounds of these detectors can be achieved.

Alternatively, we consider a further configuration, indicated as IAXO$_\textrm{r}$, with 
optimized energy resolution.
Though generally realistic, we nevertheless note that the energy resolution in this configuration is non-trivial 
and may require new technologies such as microcalorimeters operated at mK temperatures~\cite{Unger2021,osti_21266472}.
Magnetic microcalorimeters (MMCs) are also studied in the IAXO collaboration for precision measurements of the axion spectrum in the energy range $<$10$\,$keV. First measurements have shown an energy resolution of 6.1$\,$eV (FWHM) at the 5.9$\,$keV $^{55}$Fe-peak~\cite{Unger2021}.
To achieve the sufficient energy resolution,
advanced cryogenic setups operating in the milli-Kelvin regime will need to be implemented and a sufficiently low background still needs to be established. We therefore have allowed for a somewhat larger background rate.

For IAXO+ we anticipate also additional improvements of the detection system combined with the enhanced magnetic field, area etc.~outlined in~\cite{IAXO:2019mpb}. 

\subsection{Results (massless axions)}
\label{sec:massless}
For our sensitivity study in the coupling space spanned by axion-photon $g_{a\gamma}$ and effective axion-nucleon coupling $g^{\text{eff}}_{aN}$, 
we are going to assume that the axion is very light ($\lesssim$20\,meV) or massless. We will briefly comment on the effects of the mass later, in Sec.~\ref{sec:sensitivity_massive}. For the moment, let us nevertheless note that in case of a discovery of an axion with a non-vanishing mass one can employ a suitable amount of buffer gas so that further measurements such as the one of the nucleon coupling can essentially be done as in the case of vanishing mass.

In the massless case, the conversion probability of an axion to a photon $P_{a \rightarrow \gamma}$ is energy independent, 
\begin{equation}
	P_{a \rightarrow \gamma}=\frac{g_{a\gamma}^2 B^2 L^2}{4}, \label{eq:conversionmassless}
\end{equation}
where $B$ is the magnetic field and $L$ the length of the conversion volume.  Other experimental parameters that we used in the calculation include the cross-section of the magnet bores $A$, the total observation time $t$, optics and detector efficiencies, $\epsilon_o$ and $\epsilon_d$, the size of the focal spot $a$, the background level $b$ and the relative energy resolution at 14.4\,keV $r_\omega$. 

We have considered several different combinations of parameters which are listed in Tab.~\ref{tab:setups}. 
The benchmark values for the BabyIAXO, IAXO and IAXO+ magnets were taken from the most recent IAXO publication~\cite{IAXO:2019mpb}. 

Because the FWHM of the Doppler-broadened iron peak is only $\sim5$\,eV, it can be assumed that the whole signal is always in one energy bin.\footnote{This is only approximately true for the configurations IAXO$_{\rm r}$ and IAXO$^+_{\rm r}$ in Tab.~\ref{tab:setups} for which we assumed an energy resolution equal to the FWHM of the emission peak.} We can therefore calculate the expected number of signal events $\mu_{\text{signal}}$ directly from the total flux given in Eq.~\eqref{eq:totalflux}
\begin{align}
    \mu_{\text{signal}} &= \Phi_a\ P_{a \rightarrow \gamma}\  A\ t\  \epsilon_o\ \epsilon_d  \\
    &\propto (g_{a\gamma}g_{aN}^{{\rm eff}})^2.
\end{align}
As already discussed at the beginning of the section, to find the $^{57}$Fe detection sensitivity, we also have to take all possible background sources into account. First, the detector background, which is quantified by the background level $b$ and which can be measured accurately at times when the magnet bores are not pointed at the sun. Second, the tail of the Primakoff spectrum, which may act as an additional background. The expected background events $\mu_{\rm back}$ are therefore given by
\begin{align}
    \mu_{\rm back}=& \int_{E^*-\frac{\Delta E_d}{2}}^{E^*+\frac{\Delta E_d}{2}}\left( \frac{\textrm{d}\Phi_a^{\rm P}}{\textrm{d}\omega}  \  \epsilon_o \epsilon_d\right)\textrm{d}\omega\ P_{a \rightarrow \gamma} A t  \nonumber\\& + b a t \Delta E_d  \\
    \simeq & \left(g_{a\gamma}^4\kappa  \epsilon_o \epsilon_d  +b a \right)\Delta E_d t \;. \label{eq:kappa}
\end{align}
The Primakoff flux $\Phi_a^{\rm P}$ as well as the efficiencies $\epsilon_o$ and $\epsilon_d$ are in general functions of $\omega$. In case of a sufficiently small energy resolution $\Delta E_d$, we can average over the energy and describe the Primakoff background using the constant $\kappa$, as done in Eq.\eqref{eq:kappa}.  
If the Primakoff background to the $^{57}$Fe-peak is detectable at 14.4\,keV, it is clear that there will be a much stronger Primakoff signal at smaller energies and we will have measured $g_{a\gamma}$ very precisely. Therefore, we either know the expected contribution from Primakoff axions to the number of background events $\mu_{\text{back}}$ or it is negligible compared to the intrinsic detector background. 

\begin{figure}[t]
    \centering
    \includegraphics[width=1\linewidth]{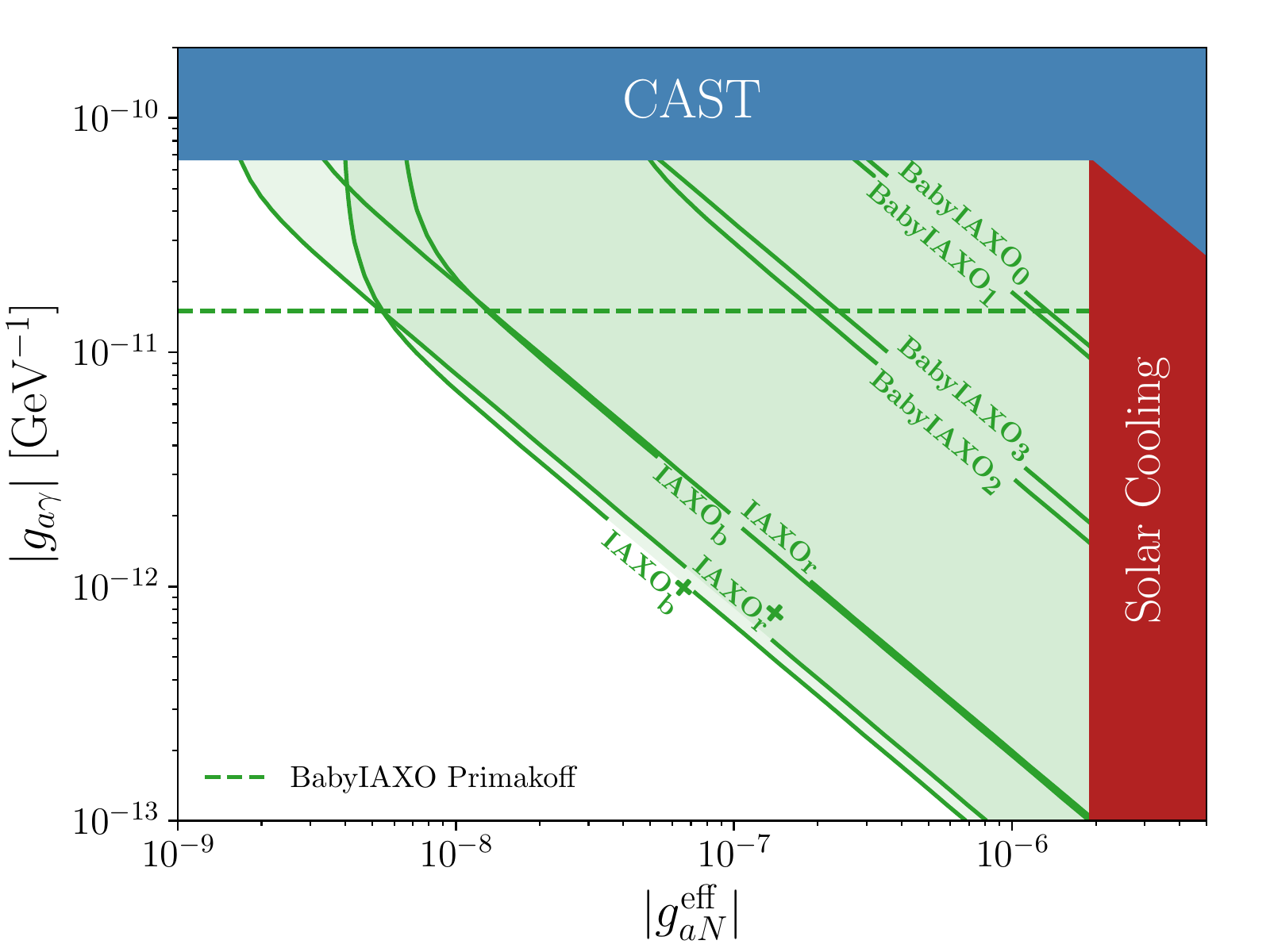}
    \caption{Model independent prediction for the  sensitivity to the axion couplings for light axions ($\lesssim20$\,meV). 
    The different regions refer to the setups presented in Tab.~\ref{tab:setups}. 
    The dark red region is the solar bound discussed in the text (cf.~Eq.~\eqref{eq:solar_bound}). 
    The dark blue region represents the latest CAST exclusion 
    regions from searches for the Primakoff flux~\cite{CAST:2017uph} and the \fe peak~\cite{CAST:2009jdc}, which we have rescaled to match the updated axion flux from \fe transitions. The dashed horizontal green line indicates the expected  sensitivity to the pure Primakoff flux. The supernova limit, Eq.~\eqref{eq:SNbound}, $g^{\rm eff}_{aN}\lesssim 1.1\times 10^{-9}$ is not shown.} 
    \label{fig:sensitivity}
\end{figure}

The $p$-value of the Poisson-distributed observed number of counts $k$ in the signal bin is given by 
\begin{align}
p=&\sum_{n=k}^{\infty} \frac{\mu_{\textrm{back}}^ne^{-\mu_{\textrm{back}}}}{n!}\ ,
\end{align}
where $k\in \mathbb{N}$. In order to find the sensitivity in parameter space, we have to calculate the expectation value of $p$ assuming a Poisson distribution for $k$ with mean $\mu = \mu_{\textrm{back}} + \mu_{\textrm{signal}}$ for each possible value of the two couplings. Note that $\mu_{\textrm{back}}$ is the sum of the usual detector background and Primakoff background. 
We regard the experiment as sensitive to a set of axion couplings if the expected $p$-value, $\langle p\rangle_k$, is smaller than 0.05.\footnote{A measurement with $ p=0.05$ would strictly speaking only amount to a 2$\sigma$ anomaly. Nevertheless, it is common to define the sensitivity in this way because it coincides with the expected exclusion limits in the case of a null result.} The resulting sensitivity curves are plotted in Fig.~\ref{fig:sensitivity}.


\begin{table*}[!t]
	\renewcommand{\arraystretch}{1.15}
	\centering
		\begin{tabular}{l|cccc|cc|cc}
			\textbf{} &   \multicolumn{4}{c|}{\textbf{\bb}} &\multicolumn{2}{c|}{\textbf{IAXO}}   & \multicolumn{2}{c}{\textbf{IAXO+}} \\
			\hline
			& \footnotesize{baseline} &\makecell{\footnotesize{no} \\ \footnotesize{optics}}& \makecell{\footnotesize{optimized} \\ \footnotesize{optics}} &\makecell{\footnotesize{high energy} \\ \footnotesize{resolution}}
			& \makecell{\footnotesize{low} \\ \footnotesize{background}}& \makecell{\footnotesize{high energy} \\ \footnotesize{resolution}}&\makecell{\footnotesize{low} \\ \footnotesize{background}}& \makecell{\footnotesize{high energy} \\ \footnotesize{resolution}}\\
			Label& \footnotesize{\bbc$_0$} & \footnotesize{\bbc$_1$} & \footnotesize{\bbc$_2$} & \footnotesize{\bbc$_3$} & \footnotesize{IAXO$_{\rm b}$} &\footnotesize{IAXO$_{\rm r}$}& \footnotesize{IAXO$^+_{\rm b}$}&\footnotesize{IAXO$^+_{\rm r}$}\\
			\hline \hline
			$B$\ [T] & 2 & 2  & 2 & 2 & 2.5 & 2.5 & 3.5 & 3.5 \\
		    $L$\ [m] & 10 & 10 & 10 & 10 & 20 & 20 & 22 & 22 \\
		    $A$\ [m$^2$] &  0.77 & 0.38 & 0.38 & 0.38 & 2.3 & 2.3 & 3.9 & 3.9\\
		    $t$\ [year] & 0.75 & 0.75 & 0.75 & 0.75 & 1.5 & 1.5 & 2.5 & 2.5 \\
		    $b$\ [$\frac{1}{\textrm{keV} \textrm{cm}^2 \textrm{s}}$] & $10^{-7}$  & $10^{-6}$  & $10^{-7}$ & $10^{-5}$ & $10^{-8}$ & $10^{-6}$ & $10^{-9}$ & $10^{-6}$  \\
		    $\epsilon_d$ & 0.15 & 0.9 & 0.5 & 0.99 & 0.99 & 0.99 & 0.99 & 0.99 \\
		    $\epsilon_0$ & 0.013 & 1 & 0.3 & 0.3 & 0.3 & 0.3 & 0.3  & 0.3  \\
		    $a$\ [cm$^2$] & 0.6 & 3800  &  0.3 &  0.3 & 1.2 & 1.2 & 1.2 & 1.2 \\
		    $r_\omega=\frac{\Delta E_{d}}{14.4\,{\rm keV}}$ & 0.12 & 0.12 & 0.12 & 0.02 & 0.02 & $\frac{5}{14400}$& 0.02 & $\frac{5}{14400}$ \\
		    
		\end{tabular}

			\caption{
List of experimental parameters adopted for all helioscope configurations which are considered in Figs.~\ref{fig:sensitivity} and~\ref{fig:massive}. As usual $B$ is the magnetic field of the helioscope, $L$ its length and $A$ the area. $t$ is the time that the helioscope is pointed at the sun. 
For these parameters we use values based on~\cite{IAXO:2019mpb}. 
As already mentioned below Eq.~\eqref{eq:merit} $\epsilon_{o,d}$ are the efficiencies of the optics and detector, $b$ is the spectral background rate per detector area, and $r_{\omega}$ is the relative spectral resolution of the detector.  
Setup~\bbc$_0$ 
is the baseline \bbc, \bbc$_1$ is a version without optics, \bbc$_{2,3}$ assume optics optimized for the 14.4\,keV line, with \bbc$_3$ including also a good energy resolution. 
In addition we show parameters from more advanced setups of IAXO and IAXO+.
}
		\label{tab:setups}

\end{table*}

Since we have assumed massless axions for these sensitivity estimates, we do not show the parameter space of DFSZ and M1 models (cf.~Sec.~\ref{sec:models}). These 
models require large masses --  $m_a\gtrsim 100\,$meV -- at those couplings, and 
helioscopes quickly lose sensitivity above $\sim$\,20\,meV (cf.~Fig.~\ref{fig:massive}).
This problem could be eased by filling the helioscope bore with a buffer gas~\cite{vanBibber:1988ge}.

Our analysis shows the potential of \bb to study areas 
of the parameter space, well beyond the solar bound and the region probed by CAST.
The different green shaded areas show the experimental potential for different configurations, 
summarized in Tab.~\ref{tab:setups}.
The most efficient setups for \bb are the ones with optimized optics (labeled \bbc$_{2,3}$ in Tab.~\ref{tab:setups}).
As evident from the table, these setups allows for an enormous reduction of the total background by limiting the focal spot area $a$ by $\sim$\,4  orders of magnitude with respect to the no-optics solution. 
Adopting this setting, 
\bb would be able to extend its detection potential also to regions of the parameter space below the Primakoff sensitivity, for 
$g_{aN}^{\rm eff} \gtrsim 10^{-7}$.
It is, therefore, possible, at least in principle, that \bb could discover axions through the \fe channel, before the Primakoff flux can be detected. 
If, on the other hand, axions have couplings in the green shaded area above the Primakoff sensitivity line (dashed green), 
one might have the opportunity to extract both couplings and derive information about the underlying axion model.

The more optimistic IAXO and IAXO+ configurations can explore an even larger area of parameter space. A noteworthy feature of these exclusion curves is their behaviour at different values of $g_{a\gamma}$. At values of $g_{a\gamma}\lesssim 10^{-11}\ \textrm{GeV}^{-1}$ the detector background dominates over the Primakoff background. Therefore the figure of merit in Eq.~\eqref{eq:merit} becomes
$f\propto \frac{\epsilon_{o}\epsilon_{d}\ g^2_{a\gamma}}{\sqrt{\Delta E_db a}}$. With the parameters given in Table~\ref{tab:setups}, the configurations with minimized background slightly outperform the ones with optimized energy resolution in this regime (cf.~Fig.~\ref{fig:sensitivity}). However, at $g_{a\gamma}\gtrsim 10^{-11}\ \textrm{GeV}^{-1}$ the Primakoff background starts to play a role and eventually dominates. In this regime the figure of merit is given by $f\propto  \sqrt{\frac{\epsilon_{o}\epsilon_{d} }{\Delta E_d\kappa }}$. The detector background becomes negligible and the configurations with optimized energy resolution are significantly more sensitive than the ones with minimized background. Therefore, the ideal detector for the $^{57}$Fe line crucially depends on the value of $g_{a\gamma}$. If \bb detects Primakoff axions, a detector with good energy resolution may be required to supress the Primakoff background to the $^{57}$Fe line. If, on the other hand, \bb only puts a stronger bound on $g_{a\gamma}$, the energy resolution becomes less important and the low background detectors may be advantageous.

\subsection{Effects of a finite axion mass}
\label{sec:sensitivity_massive}

On the production side, the axion mass only becomes relevant at scales of the solar temperature $\sim$\,keV. Therefore, we can safely regard the solar axion flux as independent of the mass.
However, on the detection side, a finite axion mass can cause decoherence between the photon and axion wave functions inside the magnet bores and lead to a signal suppression. 
The full expression for the conversion probability $P_{a \rightarrow \gamma}$ of axions into photons in the helioscope reads~\cite{Raffelt:1987im,vanBibber:1988ge} 
\begin{equation}
 \label{eq:conversionmassive}
    P_{a \rightarrow \gamma}=\frac{g_{a\gamma}^2 B^2 L^2}{4} \times \frac{2(1-\cos(qL))}{(qL)^2} \ ,
\end{equation}
where $q$ is the transferred momentum given by $m_a^2/(2\omega)$ and $\omega$ is the energy of the axion. 
The suppression becomes relevant for masses above $\sim$\,20\,meV, the exact value depending on the length of the magnet. 

\begin{figure}
    \centering
    \includegraphics[width=1\linewidth]{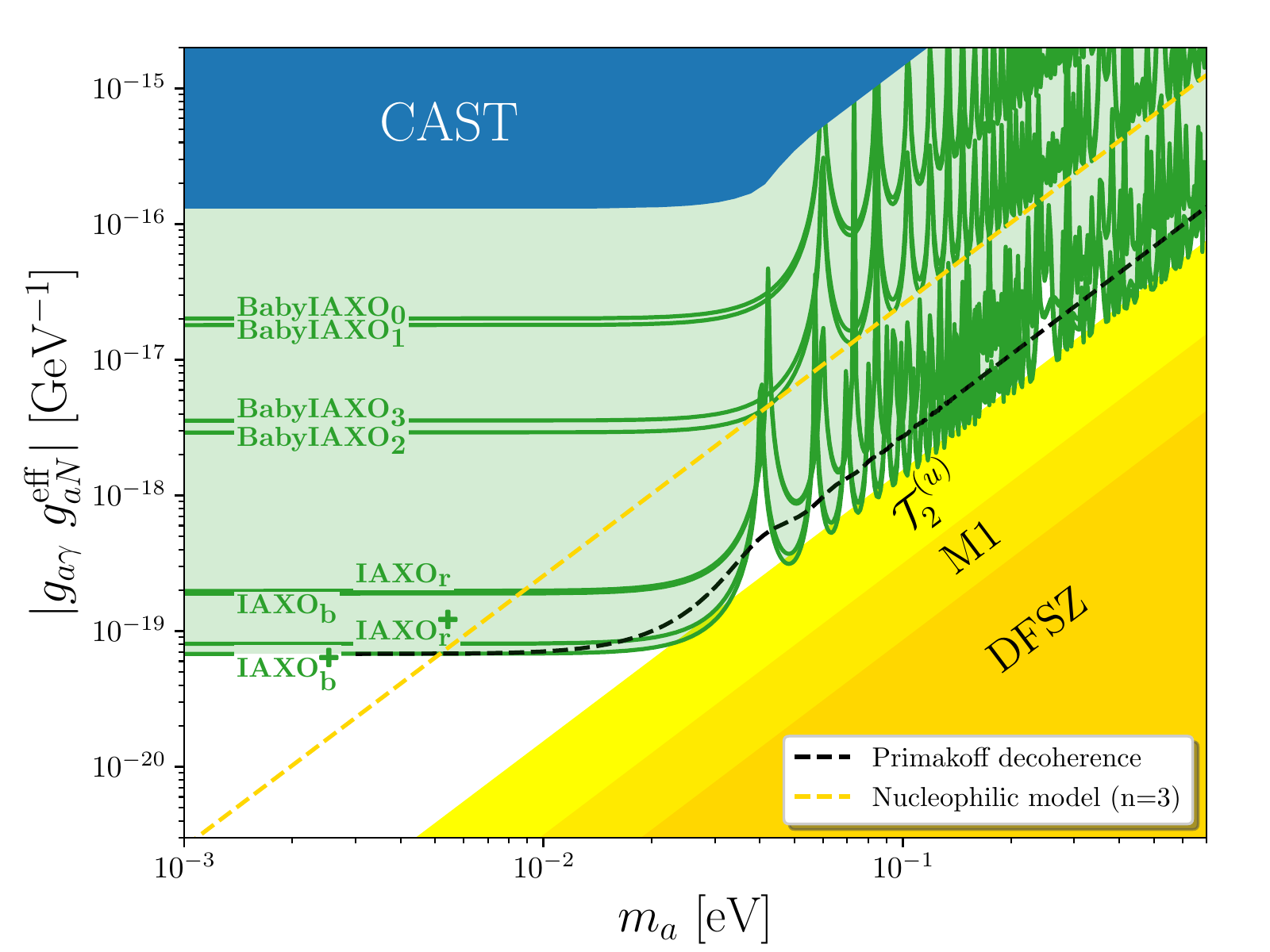}
    \caption{Model independent prediction of the IAXO sensitivity to the $^{57}$Fe peak, 
    assuming that axions are produced only through the axion coupling to nucleons. In contrast to Fig.~\ref{fig:sensitivity}, we directly show the sensitivity of the various setups to the coupling combination $g_{a\gamma} g_{aN}^\textrm{eff}$ under the assumption that the Primakoff background is negligible.
    The oscillations at higher masses are due to the form factor in the conversion probability.
    For comparison we show the effect of decoherence for a Primakoff spectrum as a dashed black line.
    The dark blue region represents the rescaled 
    CAST result~\cite{CAST:2009jdc}. 
    The dark yellow region indicates the parameter space expected for the DFSZ model. In brighter shades of yellow, we show the flavor non-universal DFSZ models M1 and $\mathcal{T}_2^{(u)}$.
    The dashed yellow 
    line, on the other hand, shows the 
    expected coupling for a nucleophilic QCD axion model of the kind presented in Ref.~\cite{Darme:2020gyx}, with $n=3$. 
All models with $n>3$ would be already accessible to \bbc.
Note that the experimental sensitivity estimates here do not assume the use of a buffer gas, which would extend the sensitivity to higher masses.}
    \label{fig:massive}
\end{figure}

The decoherence effect can be compensated at the cost of some absorption by feeding a buffer gas into the bores~\cite{vanBibber:1988ge,CAST:2008ixs}. 
This is why the sensitivity study to effectively massless (i.e.\ $m_a \lesssim 20\ \mathrm{meV}$) axions in the previous section serves as a good benchmark.
Nevertheless, we also want to explicitly investigate the sensitivity 
to massive axions without a buffer gas. 
To do this, we assume that the background from Primakoff axions is negligible and that instead the detector background dominates. In this case, the background is independent of any axion properties and the signal depends on the product of the two couplings, $g_{a\gamma} g^\mathrm{eff}_{aN}$, as well as the mass. 
The statistical analysis is equivalent to the one in the massless case. 
From the expected signal and background we calculate the expected $p$-value and draw the sensitivity curves where $p$ is expected to be smaller than 0.05. 
The results for a selection of viable setups are plotted in Fig.~\ref{fig:massive}. 
The regions shaded in yellow indicate the coupling relations for the DFSZ, M1 and $\mathcal{T}_2^{(u)}$ models (cf.~Sec.~\ref{sec:models}).

The sensitivity curves are very similar to typical helioscope exclusion plots in the coupling vs.\ mass plane with two noteworthy exceptions. Because the $^{57}$Fe line is highly energetic at 14.4\,keV, the transferred momentum $q$ is smaller than for axions of the same mass from other solar processes. As a result the decoherence effect only becomes relevant at slightly higher masses in comparison to -- for instance -- Primakoff axions. To illustrate this effect, we have plotted the expected sensitivity of the IAXO$^+_{\rm b}$ setup with a decoherence factor from a Primakoff spectrum as a dashed black line in Fig.~\ref{fig:massive}. Furthermore, the oscillations of the form factor for large $qL$ are clearly visible in the $^{57}$Fe exclusion lines while they are washed out in the case of the broadband Primakoff spectrum. 

\section{Discussion and conclusion}
\label{sec:Conclusion}

In this work, we have presented the first dedicated investigation of the \bb and IAXO potential to detect 14.4\,keV axions from \fe transitions in the sun.
The analysis is based on a recent reevaluation of the matrix elements for the 
nuclear transition~\cite{Avignone:2017ylv} and an updated solar model~\cite{Vinyoles:2016djt}. 
We have carefully considered different realistic setups for the detectors and optics
to evaluate which combination gives the best performance for 14.4\,keV axions. 
The configuration parameters are listed in Tab.~\ref{tab:setups}.

Our results, summarized in Fig.~\ref{fig:sensitivity} and~\ref{fig:massive}, show that already \bb will be able to study a large section of interesting axion parameter space, well beyond the region accessed by CAST, particularly 
in the configurations with an optimized X-ray optics. 
The potential will be greatly improved with IAXO and IAXO+.

The figures also show representative QCD axion models. 
The yellow band in Fig.~\ref{fig:massive} represents the parameter area spanned by generalized DFSZ models~\cite{DiLuzio:2017ogq,DiLuzio:2021ysg} (QCD axion models with two Higgs doublets plus a singlet scalar field, allowing for flavor non-universality).
There, we present the specific examples of the classical DFSZ axion model, and of the M1 and $\mathcal{T}_2^{(u)}$ models discussed in Sec.~\ref{sec:models}.
We are not showing the well known KSVZ axion model since
an accidental cancellation reduces its effective coupling to nucleons relevant in the \fe transition (see Sec.~\ref{sec:models}).

Although 
\bb is expected to have enough sensitivity to 
explore large sections of the parameter space for these models~\cite{DiLuzio:2021ysg},
a sizable 
axion flux from \fe transitions requires large axion masses, 
where \bb loses sensitivity.
This problem can be eased with the use of a buffer gas~\cite{vanBibber:1988ge}, a technique already tested in CAST~\cite{CAST:2017uph}.
The sensitivities shown in the figures do not account for this option in \bb nor in its scaled up versions. 
A dedicated study may show if a buffer gas may allow to probe the M1 or other DFSZ-like models through the \fe line in the near future. 

Less minimal models for QCD axions may present larger couplings to nucleons and be better accessible through the \fe channel. 
For example, the nucleophilic QCD axion models presented in Ref.~\cite{Darme:2020gyx} (and shown in Fig.~\ref{fig:massive}), have exponentially large couplings to nucleons\footnote{In model A of Ref.~\cite{Darme:2020gyx}, 
one has
\begin{align}
    g_{a\gamma}g_{aN}^{\rm eff}\sim 
    \frac{2^{2n}\alpha_{\rm em}}{2\pi\,f_a}
    \frac{m_n}{f_a}
\sim 2^{2n+2}\cdot 10^{-17}\,
\left(\frac{m_a}{\rm eV}\right)^2    \,{\rm GeV^{-1}} 
\end{align}
where $n+1$ is the number of Higgs doublets in the model. 
The parameter $n$ is constrained to $n\lesssim 50$ 
in order to avoid sub-Planckian Landau poles \cite{DiLuzio:2017pfr}. 
}
and are efficiently produced in \fe transitions even at lower axion mass.
As shown in Fig.~\ref{fig:massive}, practically, the entire class of these models will be accessible already to \bbc, even without the need for a buffer gas. 

One should nevertheless keep in mind that most of the region shown in the figures is in tension with astrophysical considerations, in particular, SN1987A (cf.~Eq.~\eqref{eq:SNbound}). As these are affected by their own uncertainties (e.g. relying on a single supernova event) it would nevertheless be comforting to have independent confirmation in more controlled setups. 
In addition, the IAXO+ setup shown in Fig.~\ref{fig:sensitivity} approaches a level of sensitivity comparable to Eq.~\eqref{eq:SNbound}. This shows a pathway to pushing beyond the astrophysical limits.

In conclusion, helioscopes of the next generation may offer a unique chance to probe an interesting range of the $g_{aN}^{\rm eff}$-$g_{a\gamma}$ parameter space. 
While this potential is expected to be greatly improved with its scaled up versions, IAXO and IAXO+, 
already \bb will have enough sensitivity to detect axions with couplings to nucleons over an order of magnitude below the solar bound (see, Fig.~\ref{fig:sensitivity}).
Furthermore, our analysis shows that already \bbc, especially if equipped with optimized optics, 
has the potential to detect through the \fe channel axions too weakly coupled to photons to give a sizable  Primakoff flux (region below the dashed green line in Fig.~\ref{fig:sensitivity}). 
In a more likely scenario, a detection of axions through \fe will be accompanied by a (larger) signal from Primakoff axions,
allowing to extract important information about its couplings to both photons and nucleons. 

Ending on an optimistic outlook, we note that discovery of an axion and its nucleon-coupling induced lines could perhaps also shed light on properties of the sun.\footnote{See~\cite{Jaeckel:2019xpa} for an opportunity to use axions with electron couplings to measure the solar metallicity.} 
For example, the strong temperature dependence may make this a good way to measure the sun's core temperature.

\section*{Acknowledgements}
We would like to thank José Crespo López-Urrutia for helpful comments and discussions. LT is funded by the Graduiertenkolleg \textit{Particle physics beyond the Standard Model} (GRK 1940).
The work of M.G.~is supported by funding from a grant provided by the Fulbright U.S.~Scholar Program and by a grant from the Fundación Bancaria Ibercaja y Fundación CAI. 
M.G.~thanks the Departamento de Física Teórica and the Centro de Astropartículas y Física de Altas Energías (CAPA) of the Universidad de Zaragoza for hospitality during the completion of this work. I.G.I and J. G. acknowledge support from the European Research Council (ERC) under the European Union’s Horizon 2020
research and innovation programme, grant agreement ERC-2017-AdG788781 (IAXO+), as
well as the Spanish Agencia Estatal de Investigación under grant PID2019-108122GB.
J.J.~is grateful to be part of the ITN HIDDeN which is supported within the European Union’s Horizon 2020 research and innovation programme grant agreement No 860881.
Part of this work was performed under the auspices of the US Department of Energy by Lawrence Livermore National Laboratory under Contract No. DE-AC52-07NA27344.
\bigskip
\appendix

\onecolumngrid

\begin{table}
	\renewcommand{\arraystretch}{1.5}
	\centering
		\begin{tabular}{l||c|c|c|c|c}
			& $^\textbf{57}$\textbf{Fe}	& $^\textbf{83}$\textbf{Kr}	& $^\textbf{169}$\textbf{Tm} &  $^\textbf{187}$\textbf{Os}	& $^\textbf{201}$\textbf{Hg}  \\
			\hline
			\hline
			$E^*$ [keV] & 14.4 & 9.4 & 8.4 & 9.7 & 1.6\\
		    $J_0$ & 1/2 & 9/2 & 1/2 & 1/2 & 3/2 \\
		    $J_1$ & 3/2 & 7/2 & 3/2 & 3/2 & 1/2 \\
		    $\tau_0$ [ns] & 141 & 212 & 5.9 & 3.4 & 144\\
		    $\alpha$ & 8.56 & 17.09 & 285 & 264 & 47000 \\
		    $\epsilon$ & $10^{-4.5}$ & $10^{-8.75}$ & $10^{-11.9}$ & $10^{-10.6}$ & $10^{-10.83}$ \\
		    $a$ [\%] & 2.14 & 11.55 & 100 & 1.6 & 13.2 \\
		    \hline
		    \makecell{$\mathcal{N}_a(r=0)$ \\ \footnotesize{ [relative to $^{57}$Fe]}} & \makebox[\widthof{\num{1.8e-3}}][c]{1}
		    & \num{1.8e-3} & \num{1.3e-4} & \num{3.0e-5} & \num{1.9e-6} 
		\end{tabular}
			\caption{Isotopes with a nuclear M1 transition and $E^*< 20$\,keV. The element abundances $\epsilon$ are taken from Ref.~\cite{Asplund}. All other values are tabled in the appendix of Ref.~\cite{Roehlsberger:2004}. The values in the last row were calculated by evaluating Eqs.~\eqref{eq:axionspersolarmatter_app} and~\eqref{eq:occupation_number} with the solar core temperature $T(r=0)=1.33$\ keV.} 
		\label{tab:isotopes}
\end{table}
\hspace{-1cm}
\twocolumngrid

\section{Axions from other nuclei}
\label{app:lines}
The phenomenological discussion in this work centers around the 14.4\,keV line of $^{57}$Fe because it is expected to give the strongest signal. In order to ensure that this is true, especially considering that the $^{57}$Fe-line suffers from strong thermal suppression, we systematically searched for alternative nuclear M1 transitions which may also generate a line in the solar axion flux.

A list of potential candidates is provided in the appendix of Ref.~\cite{Roehlsberger:2004} in form of a list of isotopes featuring low-energy nuclear transitions. All of the calculations in Sect.~\ref{sec:ALPs_from_nuclear_transitions} equally apply to M1 transitions of nuclei other than $^{57}$Fe. Hence, we only need to compare their respective axion flux per mass. This is expressed in Eq.~\eqref{eq:axionspersolarmatter} above as
\begin{equation}
    \mathcal{N}_a=\mathcal{N}\omega_1 \frac{1}{\tau_0}\frac{1}{1+\alpha} \frac{\Gamma_a}{\Gamma_\gamma}\ .
    \label{eq:axionspersolarmatter_app}
\end{equation}
The nuclear matrix elements entering in the ratio of axion to photon emissions have not been computed with equal precision for the various nuclear transitions. It is however reasonable to assume values of order 1 for the dimensionless constants $\beta$ and $\eta$. Furthermore, the E2/M1 mixing ratio is already close to its ideal value of zero for the $^{57}$Fe line. The factor $\frac{\Gamma_a}{\Gamma_\gamma}$ is therefore expected to be comparable (or smaller than the one of $^{57}$Fe in case of a large $\delta$) for all nuclear transitions and
we can focus on the combination of abundance, occupation number and inverse lifetime.

First sizable differences appear in the isotope's number density $\mathcal{N}$. To estimate its value for all radii in the sun, we need to make two assumptions. First that the contribution of one isotope to the total element abundance is constant throughout the sun and identical to the one found on earth, which we denote as $a$. And second that the radial density profiles of heavy elements only differ from the one of iron by a 
constant factor. These two reasonable assumptions allow us to compare the respective values of $\mathcal{N}$ by just multiplying the photospheric abundance with the isotope abundance on earth. The former is tabulated in Ref.~\cite{Asplund} on a logarithmic scale. 
We give the ratio of the number density of the element in question normalized to the one of hydrogen, i.e.\ $\epsilon \equiv N_X/N_H$, in table~\ref{tab:isotopes}.

The thermal occupation number $\omega_1$ crucially depends on the transition energy $E^*$ and can be written as~\cite{Moriyama:1995bz,CAST:2009jdc}
   \begin{equation}
       \omega_1 = \frac{(2J_1+1)e^{-E^*/T}}{(2J_0+1)+(2J_1+1)e^{-E^*/T}},
       \label{eq:occupation_number}
   \end{equation}
where $J_0$ and $J_1$ are the total angular momentum quantum numbers of the ground and excited state, respectively. 

We collected all isotopes from the appendix of Ref.~\cite{Roehlsberger:2004} with a nuclear M1 transition below 20\,keV in Tab.~\ref{tab:isotopes}. The last row indicates the size of the axion flux from these nuclear transitions relative to the one from $^{57}$Fe. It becomes clear that the strong Boltzmann suppression in the case of $^{57}$Fe is compensated by a relatively large abundance and a small internal conversion coefficient. Even though the values in Tab.~\ref{tab:isotopes} should be understood as very rough estimates, this clearly indicates that the $^{57}$Fe line generates the strongest axion signal from nuclear de-excitations.

\bibliographystyle{apsrev_mod}
\bibliography{references.bib}

\end{document}